\documentclass[10pt,twocolumn]{IEEEtran}
\usepackage{multicol}

\usepackage{color}
\usepackage{xcolor}
\usepackage{soul}
\usepackage{amsmath}
\usepackage{graphics}
\usepackage{setspace}
\usepackage{cite}
\usepackage{latexsym}
\usepackage{float}
\usepackage{epsfig}
\usepackage{multirow}
\usepackage{cite,cases,url}
\usepackage{amssymb}
\usepackage{graphicx}
\usepackage{epstopdf}
\usepackage{balance}
\usepackage{subcaption}
\usepackage{enumerate}
\usepackage{array}
\usepackage{algorithmic}
\usepackage{algorithm}
\usepackage{enumitem}
\usepackage[normalem]{ulem}
\useunder{\uline}{\ul}{}

\newcolumntype{L}{>{\centering\arraybackslash}m{5cm}}
\newcolumntype{K}{>{\centering\arraybackslash}m{6cm}}
\newcolumntype{P}{>{\centering\arraybackslash}m{2.3cm}}
\newcolumntype{M}{>{\raggedright\arraybackslash}m{2cm}}
\newcolumntype{N}{>{\raggedright\arraybackslash}m{2.5cm}}

\newcommand\blfootnote[1]{%
  \begingroup
  \renewcommand\thefootnote{}\footnote{#1}%
  \addtocounter{footnote}{-1}%
  \endgroup
}
\begin{document}

\title{UAV-Assisted Attack Prevention, Detection, and Recovery of 5G Networks}

\author{
\IEEEauthorblockN{Aly Sabri Abdalla$^{\dag}$, Keith Powell$^{\dag}$, Vuk Marojevic$^{\dag}$, and Giovanni Geraci$^{\star}$}\\ \vspace{0.2cm}
\normalsize\IEEEauthorblockA{$^{\dag}$Dept. Electrical and Computer Engineering,  Mississippi State University,
Mississippi State, MS\\
$^{\star}$Dept. Information and Communication Technologies, Universitat Pompeu Fabra, Barcelona, Spain}
}

\maketitle

\begin{abstract}
\blfootnote{
Accepted for publication in IEEE Wireless Communications
; August 2020.  }
Unmanned aerial vehicles (UAVs) are emerging as enablers for supporting many applications and services, such as precision agriculture, search and rescue, temporary network deployment or coverage extension, and security. UAVs are being considered for integration into emerging 5G networks as aerial users or network support nodes. 
We propose to leverage UAVs in 5G to assist in the prevention, detection, and recovery of attacks on 5G networks. Specifically, we consider jamming, spoofing, eavesdropping and the corresponding mitigation mechanisms that are enabled by the versatility of UAVs. We introduce the hot zone, safe zone and UAV-based secondary authorization entity, among others, to increase the resilience and confidentiality of 5G radio access networks and services. We present simulation results and discuss open issues and research directions, including the need for experimental evaluation and a research platform for prototyping and testing the proposed technologies.
\end{abstract}

\IEEEpeerreviewmaketitle

\section{Introduction}
\label{sec:intro}

Wireless communications have gone through decades of innovation. 
Cellular communications networks have provided true mobile broadband services since the deployment of the 4G long-term evolution (LTE). 5G networks will enable new 
use cases beyond mobile broadband where low latency, high reliability or massive connectivity is needed.

5G encompasses many networking principles, from base station-controlled to ad hoc networking. All of these share the common requirement that peers or network support nodes need to be trustworthy. Otherwise, a user may give away personal information to unauthorized entities, accept wrong information, receive degraded service, or waste resources.

Security encompasses a series of attributes that are required for a communications system to protect users, messages and services from being exploited, misused or blocked. Wireless communications systems are especially vulnerable because information is signaled over the air and cannot be completely shielded.

Security has not been at the foundation of wireless communications technology development, but is becoming increasingly important. 
Since 5G systems will be extensive, complex, customizable and to a large extent software defined, these systems will have a larger attack surface than 4G networks. 
They will also be more attractive to attacks since they will be providing critical services. 
Despite dedicated specifications that establish the 5G security framework, researchers have already found important deficiencies and potential security vulnerabilities~\cite{jover2019security}. 

Some of the main threats to cellular networks are founded on the trust relationship between a user equipment (UE) and the network. 
Specifically, a UE follows the instructions coming from the base station.  
Many control signals are sent in the clear and can be easily reproduced for launching spoofing attacks. 
5G allows null encryption and the network can decide to use it.
whether to use encryption or not. Hence, eavesdropping is possible not only for capturing control information, but also user data.
The network can request 5G users to provide location updates or their globally unique Subscription Permanent Identifier (SUPI); a fake base station can leverage this. 
5G can encrypt the SUPI, but there may be instances where this is not implemented~\cite{jover2019security}. 

The reliance on a number of control channels and signals enables efficient and tailored 5G network implementations. 
While the 4G signaling frame is structured, the control signals and channels can be personalized in 5G. 
Nevertheless, if these are disturbed, the system performance will greatly suffer because admission control, resource allocation, link adaptation, and so forth depend on them. 
Hence, it is important to harden 5G systems against eavesdropping, spoofing, and jamming.

While 
it is important that security  
be considered 
during the system design, as cellular standards and use cases evolve, it is impossible to foresee all threats. 
Security thus needs to be considered during the design and operation. 
In this paper, we discuss new opportunities and challenges of using unmanned aerial vehicles (UAVs) for increasing the availability and privacy 
of 5G services.

Massive deployments of UAVs require coordination through radio signaling. For example, UAVs need to report their presence and communicate with other UAVs, terrestrial networks and ground users. 
Several working groups of the Third Generation Partnership Project 
are developing the specifications for integrating UAVs into cellular networks. 
 A network-connected UAV can be a UE, a communications relay, or an aerial base station~\cite{GerGarLin2019CTN}. 
These and other standardization efforts are in response to the increasing availability of UAVs and the emerging use cases and services they will enable in many areas important to society, including emergency response, transportation, delivery, precision agriculture, and entertainment \cite{3GPP36777}.

The research community has been looking into physical layer security approaches to ensure the confidentiality and security of 5G networks \cite{YanWanGer2015} \cite{UAVsafeguard}. 
The work presented in~\cite{PHYUAV} discusses how a UAV-assisted system can be used for physical layer security enhancements of advanced cellular networks along with the different challenges that exist. 
Most of the existing works limit their focus on passive eavesdropping attacks, which is only one of many wireless system attacks.
Our contributions in this paper can be summarized as follows:
\begin{itemize}
    \item We provide a broader discussion of radio frequency (RF) attacks on 5G and beyond networks.
    \item We propose different ways of using UAVs, individually or in groups, for enhancing the 5G cellular network security in three phases: prevention, detection, and recovery from fundamental wireless attacks. 
    \item We present illustrative use cases with numerical results that show the potential benefits of the proposed mechanisms.
    \item We discuss the important open issues and research directions.
\end{itemize}

This paper is organized as follows. In Section II, we elaborate on a variety of emerging and new technologies for UAVs  
to prevent RF attacks to 5G networks, or help with the attack detection and recovery. 
Section III presents three relevant case studies with numerical results. Section IV provides an outlook on research, and Section V draws the conclusions.

\begin{table*}[ht]
\centering
\caption{UAV-assisted attack prevention, detection and recovery techniques. The bold entries are elaborated in this paper.}

{\begin{tabular}{|p{2cm}|p{4.9cm}|p{4.5cm}|p{4.9cm}|}
\hline
\textbf{Category} & \textbf{Attack Prevention} & \textbf{Attack Detection} & \textbf{Attack Recovery}
\\ \hline
\vspace{0.001 in}
Eavesdropping &
\vspace{-0.05 in}
\begin{list}{\labelitemi}{\leftmargin=0.2em}
    \item {\textbf{\small UAV artificial noise transmission, creating safe zones}}
    \item {\small Coordination of safe zones and UAV relaying: create a safe zone and use a UAV relay to transport information in and out of the safe zone}
    \vspace{-0.1 in}
\end{list}

&
 
\vspace{-0.05 in}
\begin{list}{\labelitemi}{\leftmargin=0.2em}
    \item {\small Sensing and monitoring, using UAVs with RF and non-RF sensors for eavesdropper localization: use a non-coherent energy detector for detection of local oscillator leakage from an eavesdropper's RF front-end as well as cameras, etc. }
    \vspace{-0.1 in}
\end{list}

&

\vspace{-0.05 in}
\begin{list}{\labelitemi}{\leftmargin=0.2em}
    \item {\textbf{\small UAV relaying}}    
    \item {\small Joint resource allocation and \newline UAV trajectory optimization: limit rate while in the attack region using UAVs}
    \vspace{-0.1 in}
\end{list}

\\ \hline
\vspace{0.001 in}
{Jamming} &

\vspace{-0.05 in}
\begin{list}{\labelitemi}{\leftmargin=0.2em}

    \item {\small Multipoint transmission, including UAV relaying and UAV base stations located closer to the users: instead of using one single link that can be attacked easily, provide multiple links, or hops using UAVs to minimize the effect of jamming attacks}
    \item {\small Beamforming and dynamic relaying, by repositioning UAVs proactively: the nulling beamformer can be used to attenuate transmissions from certain angles}

    \vspace{-0.1 in}

\end{list}

&

\vspace{-0.05 in}
\begin{list}{\labelitemi}{\leftmargin=0.1em}
\parindent -0.2in 
    \item {\textbf{\small RF environment sensing and \newline processing with UAVs}}
    \item {\small UAV-assisted localization of \newline jammers: monitoring high power beams with a single UAV, or a UAV swarm, can help minimize the search area and the search time because of their 3D mobility and authonomy.}
    \vspace{-0.1 in}

\end{list}
&

\vspace{-0.05 in}
\begin{list}{\labelitemi}{\leftmargin=0.2em}

    \item {\small UAV relaying}
    \item {\textbf{\small Hot zones enabled by redundant UAV points of transmission or reception}}
    \vspace{-0.1 in}

\end{list}

\\ \hline

\vspace{0.001 in}
Spoofing &

\vspace{-0.05 in}
\begin{list}{\labelitemi}{\leftmargin=0.2em}
   
    \item {\textbf{\small UAV-enabled secondary \newline authorization}}
    \item {\textbf{\small UAV-originating secondary signal sources}}

    \vspace{-0.1 in}
\end{list}
&
 
\vspace{-0.05 in}
\begin{list}{\labelitemi}{\leftmargin=0.2em}
    \item {\textbf{\small RF environment awareness with 
    distributed UAVs}}
    \item {\small Virtual channel modeling: model the channel under abnormal and normal conditions to distinguish between normal behavior and a spoofing attack}

    \vspace{-0.1 in}
\end{list}

& 
\vspace{-0.1 in}
\begin{list}{\labelitemi}{\leftmargin=0.2em}
    \item {\small Cooperative localization using UAVs: UAVs distributed to identify inconsistencies with the authenticated node information}
    \item{\small Multi-hop transmission: provide higher power focused transmissions closer to the affected nodes }
    \vspace{-0.1 in}
\end{list}

\\ \hline

\end{tabular}%
}

\label{tab:survey}
\end{table*}

\section{UAV-Assisted Attack Prevention, Detection and Recovery}
\label{sec:contribution2}

5G networks are expected to support reduced latency, significantly more users and throughput, and a high level of reliability. Therefore, it is necessary now more than ever to ensure 5G networks are safe from potential attacks. Networks must be built with prevention mechanisms in place to stop attacks from having a chance to significantly degrade the network.
Attack prevention can generally be done by (1) hardening the system to make attacks infeasible or unattractive or by (2) effectively 
taking preventive countermeasures to be able to absorb potential future attacks. 
For attacks that cannot be prevented, the first phase of system resilience is the ability to detect abnormal behavior. 

Multiple challenges exist when attempting to detect an ongoing attack. 
Once the presence of an attack is acknowledged, it is important to understand what type of attack is occurring. Once the attack type is recognized, a proper recovery procedure can be employed.

After a network attack happens and is identified, recovery mechanisms need to step in as quickly and effectively as possible to minimize the damage. 
A recovery procedure may focus on improving the network quality for users without disabling the attacker, or focus on reducing the effectiveness of an attacker, such that it no longer has a influence on the network.
Table~\ref{tab:survey} presents a summary of various prevention, detection, and recovery techniques that can be used by a UAV-assisted framework to augment the security of 5G networks. Irrespective of the 5G waveform or frequency band, securing the wireless access is critical and we therefore propose 
UAVs in different roles, agnostic to the specific 5G physical layer or service. A selection of the approaches listed in Table~\ref{tab:survey} are displayed in bold and will be discussed in further detail in this section. It is important to mention that UAVs have flight restrictions and are power-limited, in general. 
Therefore, some of the proposed techniques may be particularly suitable as temporary solutions for critical scenarios, where high-value resources must be protected.
\subsection{Eavesdropping}
Eavesdropping is a privacy and confidentiality attack. It occurs when 
personal data 
is captured illegitimately. 
Eavesdropping is only effective if the eavesdropper observes a good signal-to-noise ratio (SNR) at its receiver and the data is unencrypted, or the eavesdropper is able to decrypt it. 
Eavesdropping is a passive attack, but can be combined with an active attack 
to improve the data interception performance. 
 
{\textbf{Safe zones:}}
\begin{figure}[t]
    \centering
    {\leftmargin=-0.2em}
    \hspace{-2.5 mm}
    \includegraphics[width=0.495\textwidth]{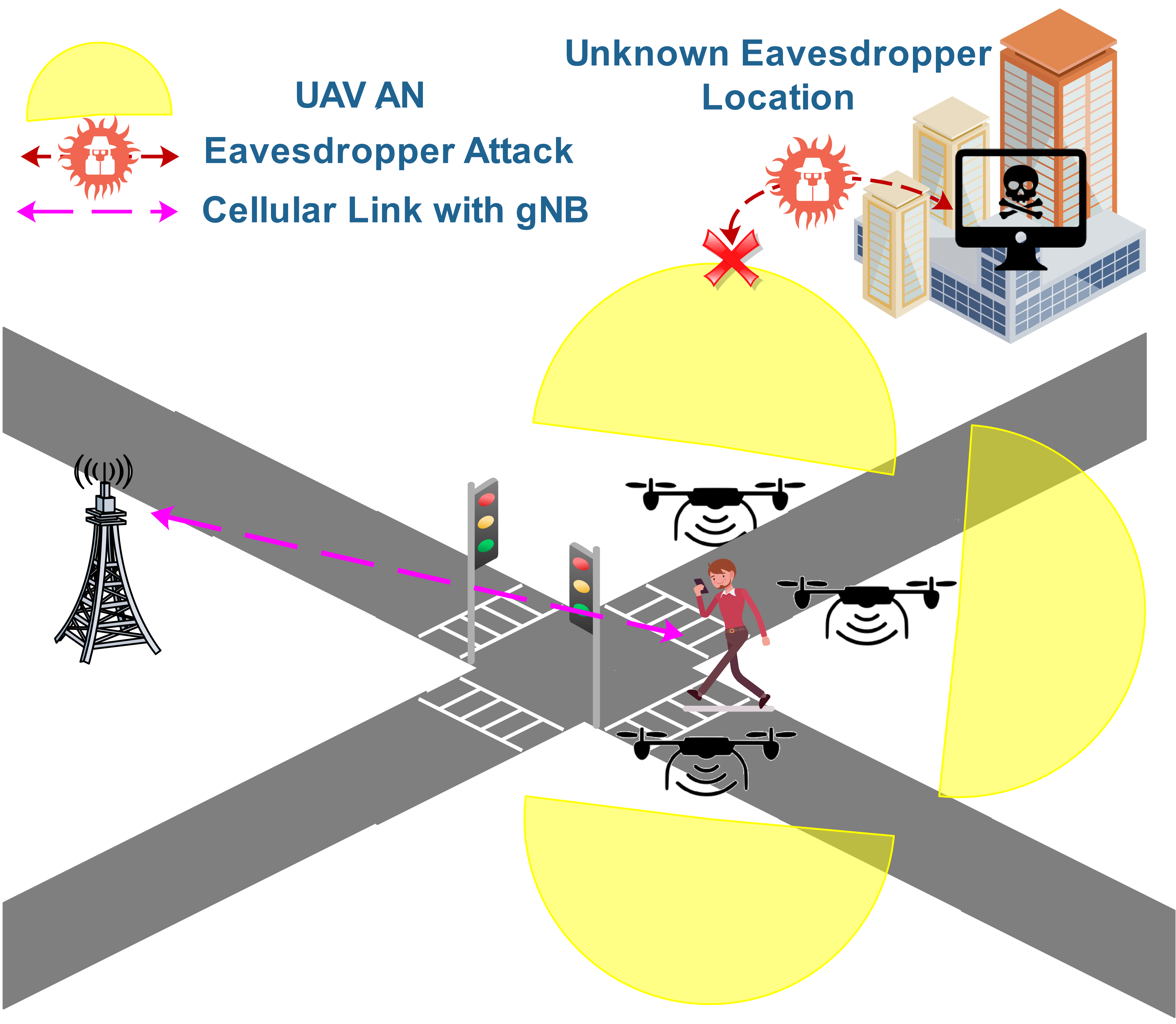}
   \centering
     \caption{Illustration of eavesdropping prevention through UAV-assisted safe zone}
    \label{fig:SafeZone}
\end{figure}
The passive nature of eavesdropping attacks means that most of these attackers will be difficult to locate, and may remain unnoticed. 
Hence, we propose an attack prevention technique that offers proactive protection for the transferred information without prior knowledge of the existence and locations of eavesdroppers. A \textit{safe zone} is a protected area around the legitimate node where a UAV or UAV swarm transmits artificial noise (AN)~\cite{UAVsafeguard} in different directions around the user to mask the user signal and prevent information leakage. This is illustrated in Fig.~\ref{fig:SafeZone}.
In such case, the geometry of the communications peers and angles of arrival/departure of the wireless link need to be 
known or discovered  
for producing AN in strategic directions.  
The beamwidth of 
the AN transmission and the number of UAVs involved will determine the level of protection against eavesdropping. 
This approach can be combined with beamforming between the base station and UE, 
or with a UAV relay discussed later. 
 
The selection of the 3D position of a UAV, 
the type of RF transmission and trajectory can potentially protect multiple ground users against eavesdropping. Safe zones can protect groups of users and the information and AN transmissions can even be coordinated to produce the best results while optimizing resources. 
For a scalable solution, terrestrial and aerial nodes of different types can be combined. 
Specifically, high altitude platforms, which can stay aloft longer and carry more payload to possibly create multiple directional beams using advanced antenna technology, should be considered.\\



\begin{figure}[h]
    \centering
    {\leftmargin=-0.2em}
    \hspace{-2.5 mm}
    \includegraphics[width=0.495\textwidth]{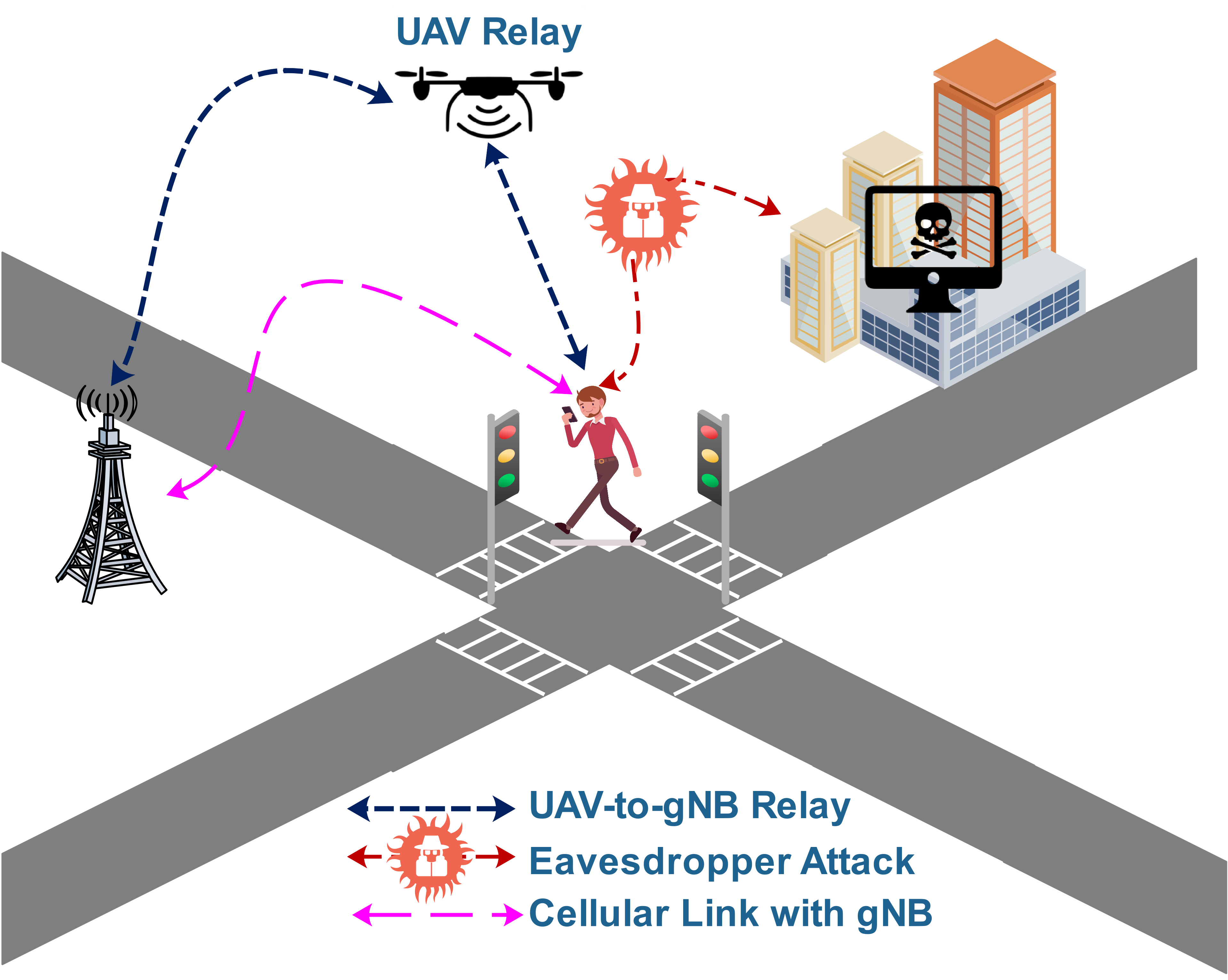}
   \centering
     \caption{Illustration of eavesdropping recovery through UAV-assisted aerial relay}
    \label{fig:Relay}
\end{figure}

{\textbf{Mobile UAV relay and resource allocation:}} UAV relays can be deployed close to the user to lower the transmission power while still maintaining a high SNR. 
Reference~\cite{PHYUAV} 
discusses the joint resource allocation and trajectory optimization to avoid eavesdroppers.
The trajectory design focuses on the position of the UAV relative to the legitimate users and eavesdroppers, whereas
resource allocation and, specifically, power control can further improve the effectiveness.  
We propose to leverage the physical environment together with the 3D mobility of UAVs and adjust the UAV speed and transmission parameters in such a way to maximize the user data rate at minimum transmission power and so minimize the information leakage.  
Together, these techniques allow the UAV to provide a strong link by focusing on the legitimate user locations while reducing the effectiveness of eavesdropping. 
This is illustrated in Fig.~\ref{fig:Relay}. 
However, in contrast to other research that assumes the eavesdropper locations are known or can be estimated, in Section III.A we show that even if the locations of the eavesdroppers are unknown, this technique can significantly lower the eavesdropping attack performance. 
\subsection{Jamming}

Jamming occurs when  
RF interference is deliberately created in order to degrade a service or prevent devices in a network to communicate with one another. As a result, a base station or set of users impacted by a jamming attack will typically experience degraded service or denial of service.

{\textbf{UAV-based jammer hunting:}}
\begin{figure}[t]
    \centering
    {\leftmargin=-0.2em}
    \hspace{-2.5 mm}
    \includegraphics[width=0.45\textwidth]{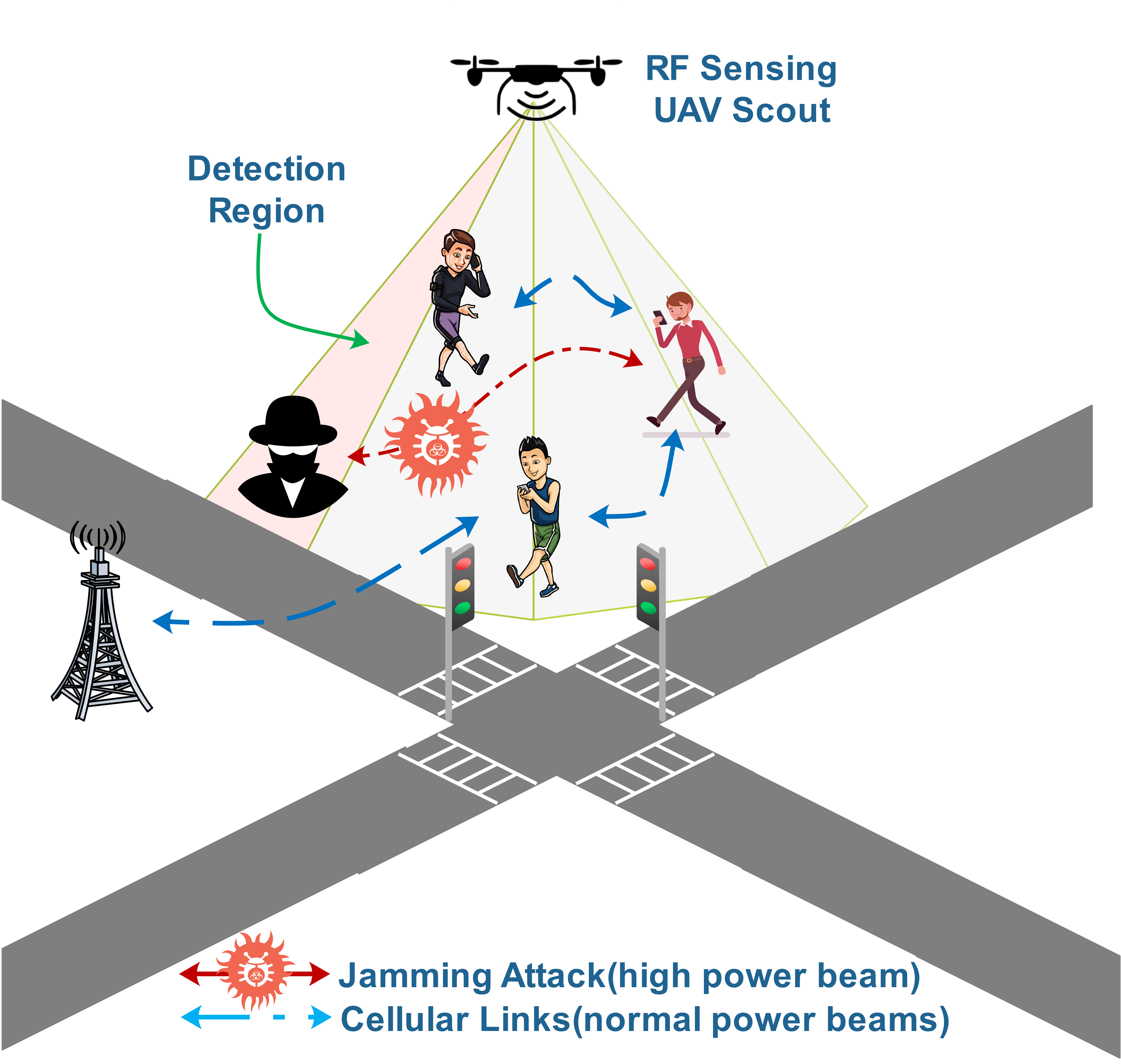}
   \centering
     \caption{Illustration of jamming detection through UAV-based jammer hunting.}
    \label{fig:JamHunt}
\end{figure}
UAVs can locate unknown signal sources much quicker than ground vehicles and without human intervention.  
Jamming, which can cause major system degradation over time while being undetected, will become less attractive if it can be detected and localized early.

The work presented in~\cite{5Gjam} proposes two reactive jamming detection mechanisms based on node collaboration.  
The first technique 
is a distributed approach where all nodes collect the performance metrics from all other connected nodes and then compare these 
to identify abnormal behavior. 
The second technique,  
uses a central node that gathers all performance metrics 
and compares these with a predefined value or threshold to identify any affected node or area. 
Both approaches can be effectively implemented with UAVs. 
UAVs can provide appropriate 3D sensing coverage 
to enhance terrestrial RF-based detection and localization of jammers and 
improve the measurement accuracy. 
UAVs can be used as pre-sensor stations equipped with powerful RF spectrum sensing capabilities for locating jammers or any high-power signal sources and obtaining their beam directionality and other transmission parameters as depicted in Fig.~\ref{fig:JamHunt}. 
A swarm of UAVs can be deployed in different directions if there is no prior information about jammers~\cite{UAVSensing}.
The collected information will be fed into a range-based or range-free method to localize the attacker. For the range-based method, the distance to the jammer is estimated with the help of 
physical measurements such as time difference of arrival, received signal strength, and time of arrival.  
The range free techniques depend on the geometric characteristics of the affected area to localize the jammer. 

The strength of using UAVs for jammer hunting is their ability to track the movement of mobile jammers. 
The change of location of a mobile jammer may trigger a new interference detection campaign using traditional methods and during this time the 
cellular network performance will suffer. 
The UAVs can tag the RF fingerprint of a jammer or use visual sensors to identify a jammer, observe its operation, and track its movement to devise and adapt the recovery mechanisms in real-time, or even predict them. 




{\textbf{Hot zones:}}
\begin{figure}[t]
    \centering
    {\leftmargin=-0.2em}
    \hspace{-2.5 mm}
    \includegraphics[width=0.495\textwidth]{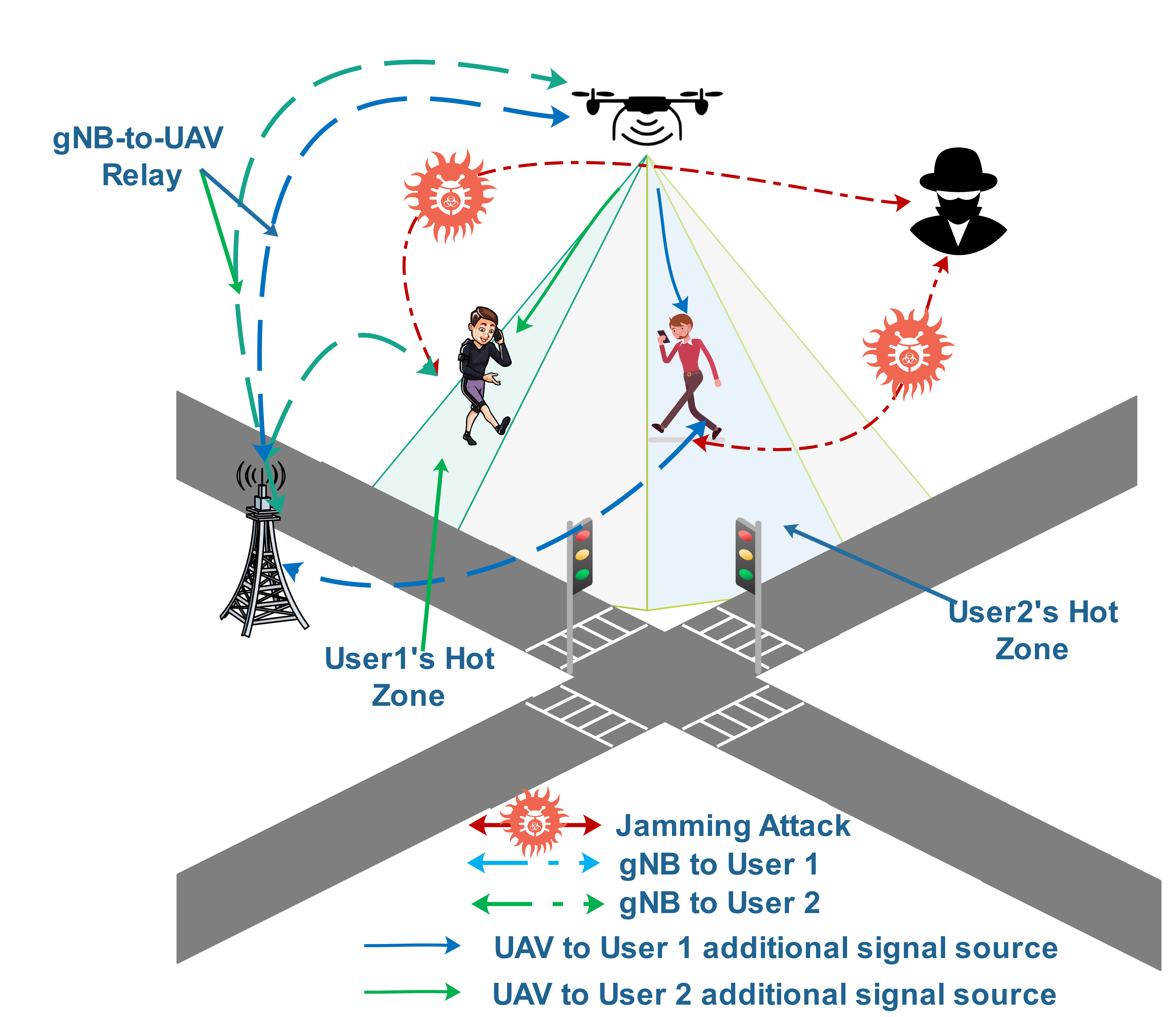}
   \centering
     \caption{Illustration of jamming recovery through UAV-based hot zone.}
    \label{fig:JamHotZone}
\end{figure}
Most anti-jamming contributions apply nulling, frequency hopping, or signal processing to minimize the effect of jamming~\cite{carrick2018method}.
These approaches can be effective in the case of static or traditional jamming attacks, but may be less 
suitable against smart or dynamic jammers. 
The technique that we propose 
minimizes the effect of jamming at the victim node through redundant transmissions from one or more deployed UAVs to create \textit{hot zones}. 
As opposed to transmitting AN 
against 
eavesdroppers, the same UAV(s) can retransmit the source signals of the legitimate transmitter 
for 
enhancing the signal strength at the legitimate receiver as illustrated in Fig.~\ref{fig:JamHotZone}. 

UAVs are suitable for multipoint transmission since they can establish 
line of sight (LoS) links to source and sink communications nodes 
and get close to the affected 
transceivers for optimal channel and signal conditions. 
The strategic 3D deployment and trajectory of UAVs can enhance the ground users' achieved rates while minimizing interference, provided tight synchronization can be established and maintained. 
Numerical results are provided in Section III.B.
A UAV, or UAV swarm, can create one or more hot zones by 
transmitting 
to one or more ground users in the region of the attack. 
This technique can be combined with  
advanced multiple-input, multiple-output, beamforming and space-division multiple access techniques to optimize and isolate resources used by terrestrial and UAV-based transmissions. 
For the case where one or more base stations are affected by strong jammers, UAV base stations can be strategically positioned and take over the radio access network functionalities. 
\subsection{Spoofing}

RF spoofing attacks occur in the form of signals or channels that are meant to deceive the network or users, disrupt communications or gain access to information or services.
GPS spoofing, in particular, occurs when an attacker provides fake coordinates or a fake time reference. 

{\textbf{UAV-based secondary authentication:}}
A single authentication mechanism and authority can be improved by adding a second form of authentication, which is already in use today for many IT services. 
As opposed to user-centric dual authentication, we propose a network-centric solution assisted by UAVs.  
It is based on dual authentication, where the computationally complex authentication mechanisms are carried out by the terrestrial 5G network and the credibility check is carried out by a UAV. 
The process begins when a new node is discovered 
by a \textit{UAV scout}. 
The scout will then start collecting the device's information, such as location, type of device, and channel state information. 
Such information can be valuable for authentication, as shown 
in~\cite{AuthCSI}, but instead of estimating the physical layer information, the UAV can capture 
and transfer it to the terrestrial 5G network and use it when the device initiates the authentication process with the network. 

{\textbf{UAV signal source:}} 
New modes of mobile communications, such as cellular-based vehicle to everything (C-V2X), heavily rely on the GPS signal as the synchronization source. 
This has been identified as a bottleneck from both the security and GPS availability standpoints. 
UAVs can mimic the global positioning system at local scale and provide accurate positioning information. 
Using standard cooperative localization mechanisms, a node may determine its location incorrectly if any of the three UAVs that is used for estimating the relative distances has been compromised.
Additional UAVs  
will provide more robustness for the location estimation as well as for identifying compromised nodes and providing them a reliable source~\cite{DroneGPSspoofing}. 

\begin{figure}[h]
    \centering
    {\leftmargin=-0.2em}
    \hspace{-2.5 mm}
    \includegraphics[width=0.495\textwidth]{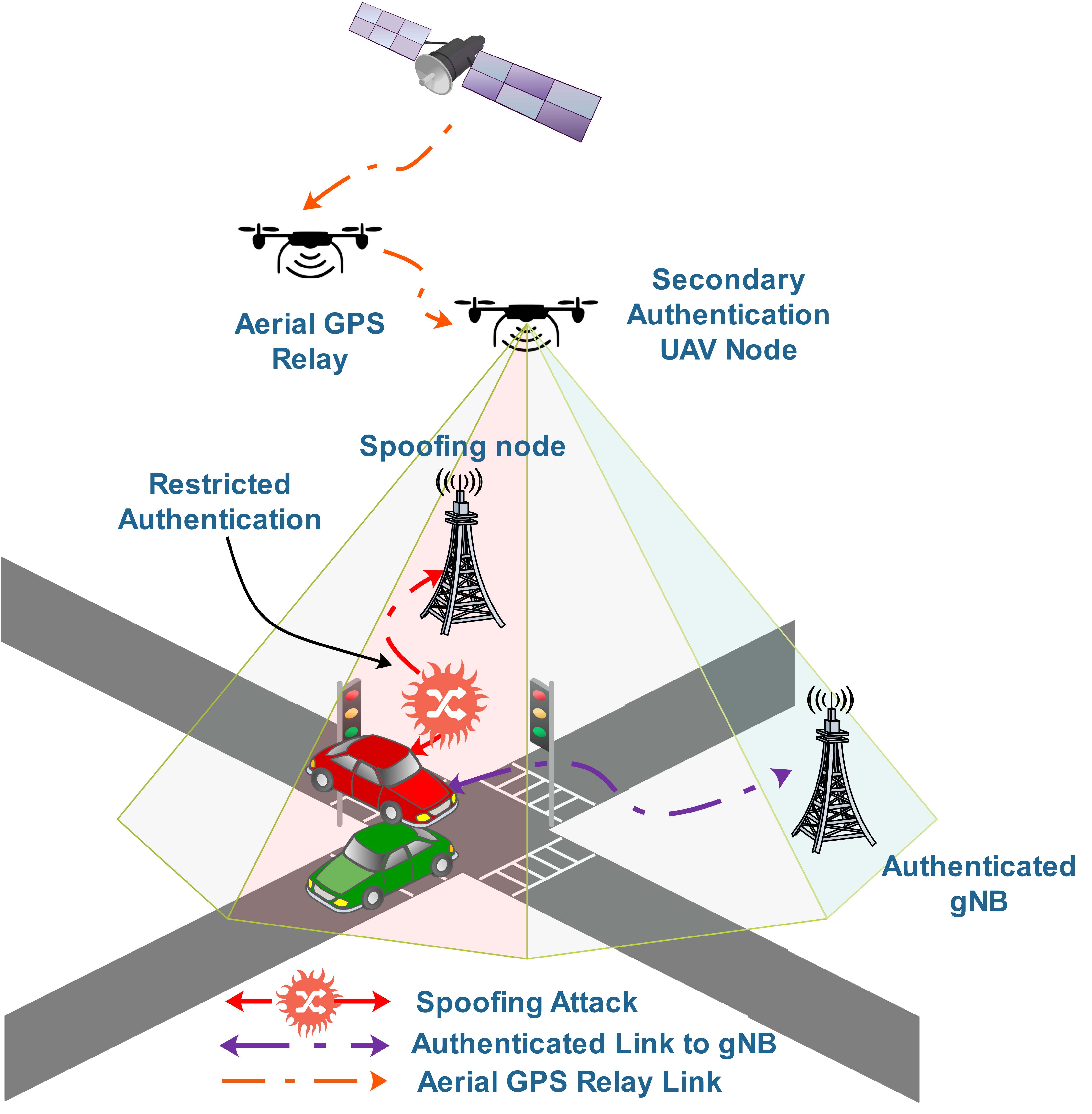}
   \centering
     \caption{UAV-assisted authentication and GPS signal source.}
    \label{fig:SpoofGPS}
\end{figure}

UAVs can also be effectively used to relay the GPS signals to network users because their mobility allows consistent LoS links with satellites as well as with 
ground users. 
A UAV relay can serve as the reference for a cluster of users that may be moving together, for instance, in a train~\cite{UAV_GPS}. Fig.~\ref{fig:SpoofGPS} illustrates the use of UAV-assisted signal source techniques to provide an authenticated GPS signal. The relayed GPS signal can be used to determine if a fake GPS signal has been transmitted. 
In addition, the  
UAV positions, trajectories and the associated channel states can be used explicitly as physical attributes for verifying traditional signal sources on the fly. 
Those aids can be thought of as a form of prevention, or detection and mitigation if an attack has already happened.
UAV-based signal sources do not need to be present all the time; they can be effectively leveraged where available and combined with other signal spoofing countermeasures, such as internal signal distribution among authenticated nodes, or verified anchors. 

{\textbf{UAV-based RF environment sensing:}}
UAVs can assist in creating awareness of the radio environment. A UAV or distributed UAV swarm collects and processes RF data as 5G signaling inspectors to monitor operation and detect any unusual RF activity in the band, such as duplicated control signals or excessive RF activity.  
These \emph{observer UAVs} will know the locations and operational parameters of legitimate network nodes and can observe (1) the general RF signaling activity and compare it to prior data as well as (2) the control signaling activity.
If the legitimate control frames, which can be dynamically configured by the 5G network, are unknown at the UAV, the UAV needs to be updated accordingly to facilitate identifying non-conforming or fake signals. 
This same technique can also be used to detect protocol-aware jammers that try to hide within one or more 5G physical channels or signals. 
Once the spoofing area is identified, it can be isolated and UAV relays deployed to 
overpower fake signals, retake control of users, or reestablish services using other resources. 

GPS spoofing attacks can be detected and mitigated through RF crowd sourcing among terrestrial and aerial nodes. 
In order to aid in this process, the legitimate nodes can exchange location and time information among themselves and the observer UAVs to identify compromised nodes~\cite{UAVspof}, inform the nodes being compromised, and provide alternative signal sources as previously described. 

\section{Case Studies}
\label{sec:case}


\subsection{Eavesdropping Attack Prevention}
\begin{figure}[t]
    \centering
    {\leftmargin=-0.2em}
    \hspace{-2.5 mm}
    \includegraphics[width=0.495\textwidth]{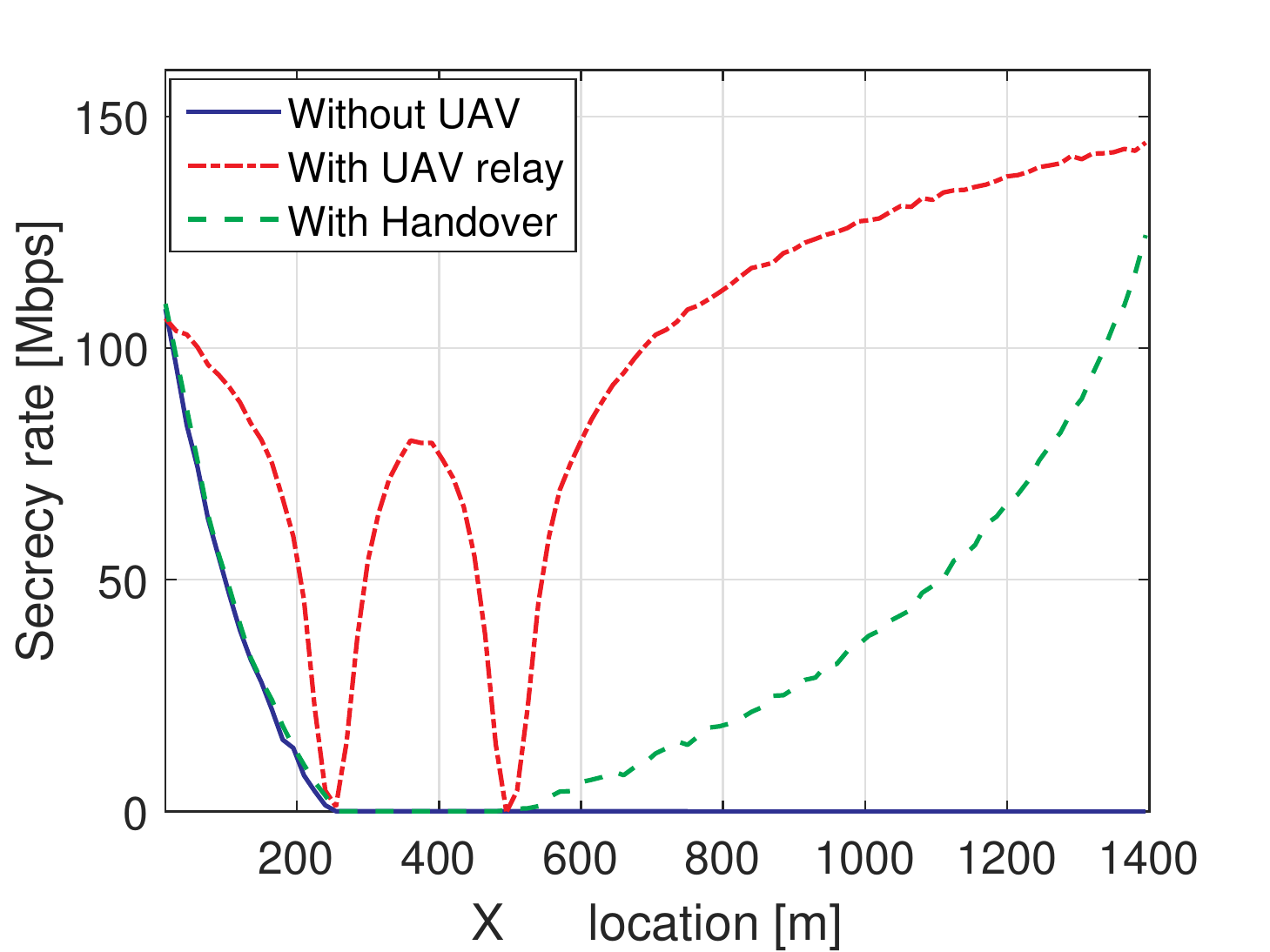}
   \centering
     \caption{Case study for eavesdropping prevention---secrecy rate achievable with an aerial relay, through base station handover, and none of them.}
    \label{fig:Secrecy}
\end{figure}
When passive attackers start listening and capturing packets exchanged between base stations and legitimate users through wiretap links, the system confidentiality and privacy are compromised. The extent to which the confidentiality of the system is compromised can be measured through the secrecy rate, the transmission rate at which no information will be decoded by the eavesdropper, and can be calculated as the difference between the legitimate and the wiretap channel capacities \cite{YanWanGer2015}. The secrecy rate is commonly used to evaluate the performance of eavesdropping mitigation techniques. Therefore, it has been adopted as a key performance indicator in our simulations.
 
We have simulated the secrecy rate of a mobile user 
that moves at constant speed in one direction and passes two eavesdroppers along its trajectory. 
A UAV is available for relaying the data from the base station to the user. The location of the 1$^{st}$ and 2$^{nd}$ base stations are at (0,0,30) and (1400,0,30) coordinates, respectively. The two eavesdroppers are located at (250,0,0) and (500,0,0), respectively. The initial position of the ground nodes is (0,0,0). The user moves on a straight line from base station 1 to 2.  
The UAV relay will follow the ground user closely at 20 m height. 
The base station power is 43 dBm while the UAV relay power is 23 dBm. The channel noise is characterized as a normal distribution with a variance of 1e-12. The downlink communications signal bandwidth is 10 MHz.
\begin{figure}[t]
    \centering
    {\leftmargin=-0.2em}
    \hspace{-2.5 mm}
    \includegraphics[width=0.495\textwidth]{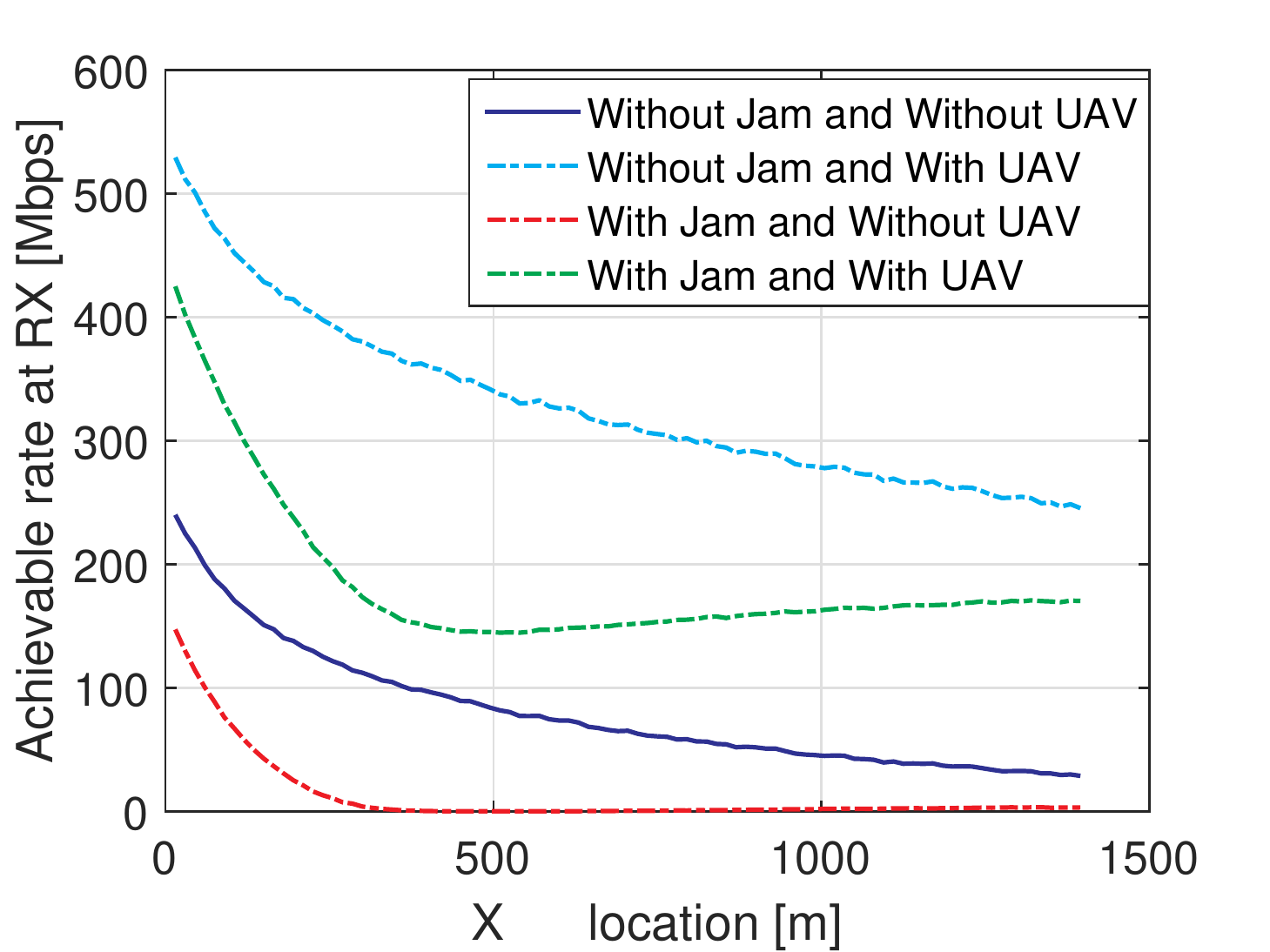}
   \centering
     \caption{Case study for jamming recovery---achievable rate with/without a ground jammer and with/without a UAV-based hot zone.}
    \label{fig:Jam}
\end{figure}
\begin{figure*}[t]
    \centering
    \includegraphics[width=0.85\textwidth]{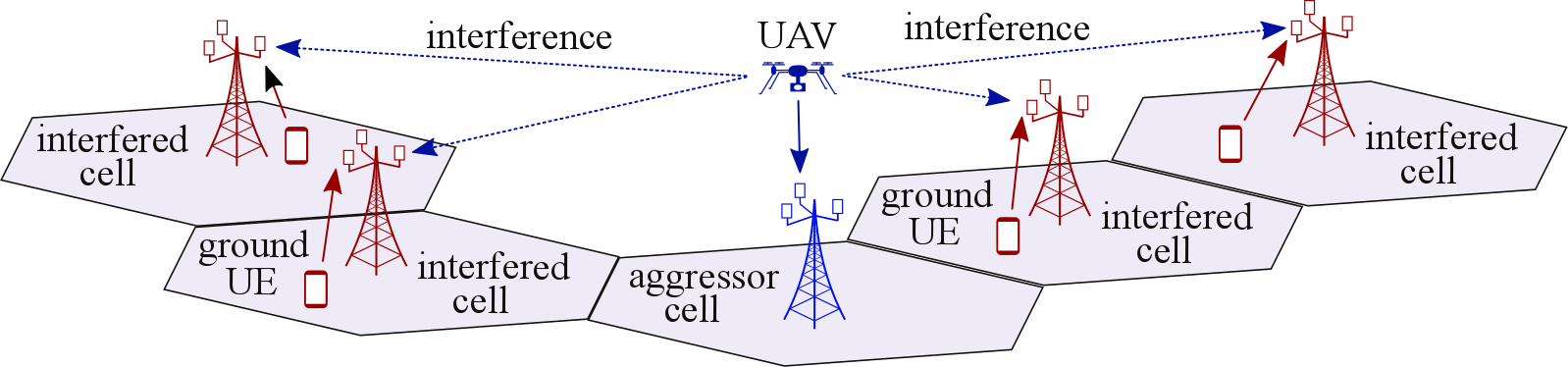}
    \centering
    \caption{UAV generating strong interference to terrestrial cells and users.}
    \label{fig:Interference}
\end{figure*}
For ground communications, we consider both distance dependent large-scale fading and small-scale fading for the channel model. We use a common air-to-ground channel model~\cite{A2G} that is composed of LoS signals, non-LoS  
signals, and multiple reflected components, which cause multi path fading, to calculate the data rates between UAV and ground receivers. The secrecy rate is calculated here for the downlink and is obtained as the difference between the achieved rate at the legitimate user from the base station, or UAV relay, and the maximum achieved rate at the two eavesdropper nodes.  
Fig.~\ref{fig:Secrecy} compares the secrecy rate at the UE for three cases: single base station transmission, single base station transmission with UAV relay, and optimal base station handover in the sense of the secrecy rate. We observe important performance improvement when using a UAV relay.

\subsection{Jamming Attack Recovery}

Jamming attacks can severely degrade the cellular network system performance and the communications capability between authenticated users. Earlier we proposed using UAVs to create hot zones as a recovery mechanism. 
The next scenario considers a moving 
ground user that is connected to one base station. The jammer schedules its resource and beam directions to attack this user, while one UAV will be deployed to create the hot zone. 
It creates another point of transmission to enhance the received signal-to-interference-plus-noise ratio (SINR) at the ground receiver while the jammer aims at degrading it. The receiver will apply equal gain combining as a receiver diversity technique to combine the transmissions from the base station and the UAV. The UAV will closely follow the movement of the ground user. 
The simulation parameters used here are equivalent to those used in Sec. III-A. The jammer location is (500,0,0) and its transmission power 15 dBW.    
The results shown in Fig.~\ref{fig:Jam} illustrate the achieved downlink rate at the UE. We observe that deploying the hot zone increases the SINR to considerably lower the effect of jamming. Also, the results show how the system starts recovery and how the rate starts to increase as the user moves away from the jammer. 
Both scenarios are scalable to multiple UEs with a single or multiple UAVs through resource sharing or advanced wireless technologies, as indicated in Section IV.

\subsection{UAV-based Attack and Recovery}

Integrating UAV devices in future cellular networks may also expose the latter to potential risks of UAV-based attacks. 
Cellular-connected UAVs may undergo radio propagation characteristics that are likely to be different from those experienced by a ground user. 
Once a UAV is flying well above cellular base stations, its uplink signal becomes more visible to multiple cells due to favorable line of sight propagation conditions~\cite{3GPP36777}. 
As a result, UAVs are capable of creating significant interference to many neighboring cells receiving uplink transmissions from their legacy ground users. This is depicted in Fig~\ref{fig:Interference}.

It is therefore critical for the cellular network to (i) detect and identify the threat posed by a UAV interfering and disrupting the service of nearby cells, and (ii) mitigate it. 
As for detection, the interfered base stations could assist the aggressor base station---i.e., the one to which the aggressor UAV is associated with---by reporting the levels of interference received on certain time-frequency resource blocks. 
Once detected, the threat can be addressed as follows: In the case of a rogue UAV, e.g., a mobile phone attached to a UAV and unauthorized to receive cellular service in the sky, the aggressor base station could stop allocating it resources and force the user to disconnect until it lowers its height. 
In the case of an authorized UAV user, the aggressor base station could still allow it to transmit while managing the interference it generates. 
This can be accomplished through inter-cell coordination, by having dedicated resources for the UAV, which will not be used by nearby cells for their ground users, or by employing UAV-specific power control, where the UAV will lower its transmit power to reduce the interference it causes.

\section{Research Directions}
\label{sec:case}

Fundamental challenges are still to be tackled for UAV-assisted network security to take off, in both the theoretical and experimental domains. In what follows, we identify what we believe are among the most compelling research directions to follow.

\begin{itemize}[leftmargin=*]

\item \textbf{5G platforms for R\&D:} The proposed UAV-assisted security enhancements need to be implemented and tested in production-like operating environments. Real data that captures diverse modes of operation, with and without attacks, need to be made available for R\&D as benchmarks for reproducing experiments and real use-cases to assess research innovations. On the technology side, this requires robust 5G platforms and configurable radios that act as attackers, relays, or enhanced 5G network nodes. On the regulatory side, it needs (i) permissions to radiate in the RF spectrum from federal spectrum authorities and license holders, and (ii) to either adhere to the established UAV flight operation rules or to be granted special permissions from federal aviation authorities.

\item \textbf{UAV experimental platforms:} UAVs have important limitations in terms of endurance, payload size and weight, and operation. Designed under the Platforms for Advanced Wireless Research program, the Aerial Experimentation and Research Platform for Advanced Wireless (AERPAW) will cover the previously mentioned technological and regulatory aspects that are needed to operate a 5G testbed with UAVs \cite{aerpaw19}. It will offer global researchers the ability to prototype and test advanced wireless technology for emerging UAV applications, giving them access to experimental base stations and UAVs that communicate via 5G and software-defined radio (SDR) networks. Modern SDRs will be deployed on the fixed towers and lightweight versions will be carried by the UAVs, both implementing advanced wireless protocols.
A number of UAVs will be available, including long endurance UAVs. The use of SDRs will allow implementing, adjusting and enhancing the proposed techniques and evaluating scalability and the impact of limited resources in real urban, suburban, and rural deployments. Advanced wireless technology, such as directional transmission and reception at the UAV and attack-aware scheduling, can be investigated with AERPAW to further improve the proposed mechanisms and effectively increase the capacity of UAVs in their envisaged roles.

\item \textbf{Data analytics for security:} Past messages, physical and environmental attributes, and information collected by multiple UAV sensors can be used to validate messages, weigh decisions, and dynamically update node and information trust metrics. Cross-discipline research is needed at the crossroads of computer vision, formal methods and communications to jointly process data from different sensors and maximize the effect of the proposed communications-centric mitigation approaches.

\end{itemize}

\section{Conclusions}
\label{sec:conclusions}
While the first 5G networks are being deployed, researchers have already identified important security threats. 
The 5G framework offers strong security mechanisms, but even if all are rigorously implemented, attackers will find a way to disturb 5G operations or gain illegitimate access to services or information. 
With the premise that 5G terrestrial networks will never be 100\% secure, we proposed to leverage UAVs for enhancing the security of 5G and beyond wireless access networks. 
UAVs are currently being considered as 5G users and network support nodes and will naturally be available for assisting terrestrial networks. 
We discussed several areas where the diversity and 3D mobility of UAVs can effectively enhance advanced wireless network security against a number of eavesdropping, jamming, and spoofing attacks, before they happen or for rapid detection and recovery. 
Simulation results showed the potential benefits of the proposed mechanisms using low-altitude aerial points of transmission. 
It is important to note that the UAV-enhanced or enabled security services need to be robust and secure to avoid introducing  
new vulnerabilities. 
Wide scale testing with advanced wireless testbeds featuring terrestrial 5G networks, SDRs and UAVs will allow prototyping and testing in order to understand the operational parameters, impact and practical limitations of UAV-assisted 5G network security. 

\section*{Acknowledgement}

A.~S.~Abdalla, K.~Powell, and V.~Marojevic are supported in part by the by NSF Platforms for Advanced Wireless Research (PAWR) program, under grant number CNS-1939334. 
The work of G.~Geraci was supported by MINECO under Project RTI2018-101040-A-I00 and by the Postdoctoral Junior Leader Fellowship Programme from ``la Caixa" Banking Foundation.

\balance

\bibliographystyle{IEEEtran}
\bibliography{main}

\section*{Biographies}
\small
\noindent
\textbf{Aly Sabri Abdalla} (asa298@msstate.edu)
is a PhD candidate in the Department of Electrical and Computer Engineering at Mississippi State University, Starkville, MS, USA. His research interests are on scheduling, congestion control and wireless security for vehicular ad-hoc and UAV networks.

\vspace{0.2cm}
\noindent
\textbf{Keith Powell} (kp1747@msstate.edu) is pursuing a PhD degree in the Department of Electrical and Computer Engineering at Mississippi State University, Starkville, MS, USA. His research interests include software radio platforms, UAV communications, and embedded systems.

\vspace{0.2cm}
\noindent
\textbf{Vuk Marojevic} (vuk.marojevic@msstate.edu) is an associate professor in electrical and computer engineering at Mississippi State University, Starkville, MS, USA. His research interests include resource management, vehicle-to-everything communications and wireless security with application to cellular communications, mission-critical networks, and unmanned aircraft systems.

\vspace{0.2cm}
\noindent
\textbf{Giovanni Geraci} (giovanni.geraci@upf.edu) is an Assistant Professor at UPF Barcelona, Spain, and was formerly a Research Scientist with Nokia Bell Labs. 
He received the IEEE PIMRC'19 Best Paper Award for his work on ``Cellular UAV-to-UAV Communications" and the 2018 IEEE ComSoc Outstanding Young Researcher Award for EMEA Region.

\end{document}